# The Lambert-W step-potential - an exactly solvable confluent hypergeometric potential

**A.M. Ishkhanyan**[1,2,3]

[1]Institute for Physical Research, NAS of Armenia, 0203 Ashtarak, Armenia
[2]Armenian State Pedagogical University, 0010 Yerevan, Armenia
[3]Institute of Physics and Technology, National Research Tomsk Polytechnic University, Tomsk 634050, Russia

We present an asymmetric step-barrier potential for which the one-dimensional stationary Schrödinger equation is exactly solved in terms of the confluent hypergeometric functions. The potential is given in terms of the Lambert $W$-function, which is an implicitly elementary function also known as the product logarithm. We present the general solution of the problem and consider the quantum reflection at transmission of a particle above this potential barrier. Compared with the abrupt-step and hyperbolic tangent potentials, which are reproduced by the Lambert potential in certain parameter and/or variable variation regions, the reflection coefficient is smaller because of the lesser steepness of the potential on the particle incidence side. Presenting the derivation of the Lambert potential we show that this is a four-parametric sub-potential of a more general five-parametric one also solvable in terms of the confluent hypergeometric functions. The latter potential, however, is a conditionally integrable one. Finally, we show that there exists one more potential the solution for which is written in terms of the derivative of a bi-confluent Heun function.

## 1. Introduction

We introduce a new exactly solvable potential for the stationary Schrödinger equation. This is an asymmetric potential step given in terms of the Lambert $W$-function which is an implicitly elementary function also known as the product logarithm [1,2]. This is one more potential for which the general solution of the problem is written in terms of the confluent hypergeometric functions. It is a four-parametric specification of a more general five-parametric potential which is also solvable in terms of the confluent hypergeometric functions. This generalized potential, however, is a conditionally integrable one.

Recently, we have presented another independent potential exactly solvable in terms of the confluent hypergeometric functions - the inverse square root potential [3]. The solution for this potential has a remarkable structure. It involves a combination with non-constant coefficients of a hypergeometric function and its derivative. In the present paper we apply another combination to derive one more exactly solvable potential, the Lambert $W$-potential.

The general solution of the Schrödinger equation for this potential is thus written as a combination of four confluent hypergeometric functions with non-constant coefficients. To treat the above-barrier reflection problem, we choose these functions such that each one of the fundamental solutions presents a separate wave moving in a certain direction. We derive a compact formula for the reflection coefficient. The structure of this formula resembles the features of both the abrupt-step and the hyperbolic tangent potentials which are reproduced by the Lambert potential in certain parameter and/or variable regions. The result shows that the reflection coefficient is smaller as compared to the two mentioned potentials because of the lesser steepness of the Lambert potential on the particle incidence side.

The approach that leads to the derivation of the Lambert and inverse square root potentials is based on the observation that if the potential is proportional to an energy-independent continuous parameter and if the potential shape does not depend on the energy, then the general Natanzon class of potentials constructed through an energy-independent transformation [4] is necessarily dropped into a few sub-potentials involving a fewer number of continuous parameters [5]. Complementary to this is the technique for construction of these new potentials using the Heun functions [6,7] developed for the quantum two-state problem in [8-10]. The technique employs a Manning-form coordinate transformation [11], and the last ingredient of the development is the application of the equations obeyed by certain functions involving the derivatives of the Heun functions [10,12]. In general, these are more complicated equations because they involve extra singularities compared with the starting Heun equations. Owing to these additional singularities, these equations suggest a wider set of covered effects. However, the singularities are *apparent* [6,7], and the solutions of these equations are written through the solutions of the Heun equations so that these equations in general do not cause additional complications compared to the Heun equations.

The inverse square root potential [3] has been derived using an equation obeyed by the derivative of the tri-confluent Heun function [6,7]. In the present paper we employ the equation obeyed by a function involving the derivative of the *bi-confluent* Heun function [6,7]. A result derived by following the outlined approach is the Lambert $W$-function asymmetric step potential. It should be noted that in fact we derive a more general five-parametric potential solvable in terms of the confluent hypergeometric functions. This potential, however, is in general a conditionally integrable one. The exactly solvable sub-potential of this potential is just the four-parametric Lambert potential that we present here. Since the derivation is itself of methodological interest, we separate that into an Appendix A.

## 2. The Lambert product logarithm potential

The Lambert $W$-function is an implicitly elementary function $W = W(z)$ that resolves the equation $W \exp(W) = z$ [1,2]. Our result is that the one-dimensional Schrödinger equation for a particle of mass $m$ and energy $E$:

$$\frac{d^2\psi}{dx^2} + \frac{2m}{\hbar^2}(E - V(x))\psi = 0, \tag{1}$$

for the potential

$$V = \frac{V_0}{1 + W(e^{-x/\sigma})}, \tag{2}$$

where $W$ is the Lambert function is exactly solved in terms of the confluent hypergeometric functions. Since the Lambert function is also known as the product logarithm, we refer to this potential as the Lambert product-logarithm potential. This is an asymmetric step-potential of height $V_0$ the steepness and asymmetry of which are controlled by the parameter $\sigma$ (Fig.1).

The general solution of the problem for arbitrary $V_0$ and $\sigma$ is written as

$$\psi(x) = z^{i\delta/2} e^{-isz/2} \left( \frac{du(z)}{dz} - i\frac{\delta + s}{2} u(z) \right), \quad z = W(e^{-x/\sigma}), \tag{3}$$

where $u(z)$ is the general solution of the scaled confluent hypergeometric equation

$$u''(z) + \left( \frac{i\delta}{z} - is \right) u'(z) + \frac{as}{z} u(z) = 0 \tag{4}$$

and the involved parameters are given as

$$a = \frac{\delta(\delta + s)}{2s} + \frac{\sigma\sqrt{m}V_0}{\sqrt{2E}\hbar}, \quad \delta = 2\sigma\sqrt{\frac{2m(E - V_0)}{\hbar^2}}, \quad s = 2\sigma\sqrt{\frac{2mE}{\hbar^2}}. \tag{5}$$

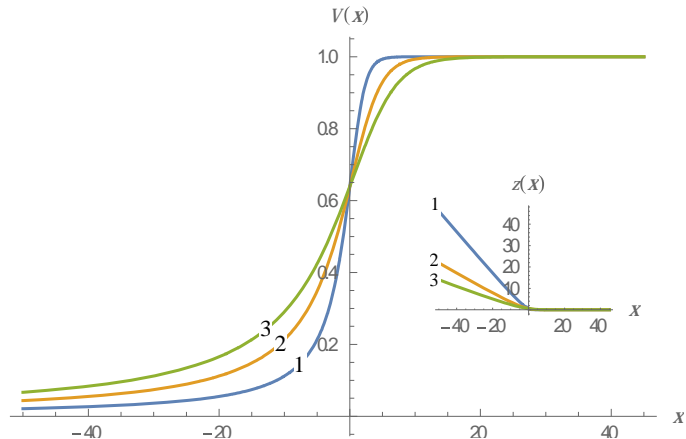

Fig. 1. The Lambert potential (2), $V_0 = 1$, $\sigma = 1, 2, 3$. The inset presents the transformation $z = W(e^{-x/\sigma})$.

Since the Lambert product logarithm barrier (2) is an asymmetric step potential, it is interesting to discuss the role of the asymmetry in the above-barrier reflection at transmission of a quantum particle above such a potential. To treat this problem, it is convenient to choose the general solution of the confluent hypergeometric equation (4) as

$$u = c_1 (isz)^{1-i\delta} {}_1F_1\left(1+i(a-\delta); 2-i\delta; isz\right) + c_2 U\left(ia; i\delta; isz\right), \qquad (6)$$

where $c_{1,2}$ are arbitrary constants and ${}_1F_1$ and $U$ are the Kummer and Tricomi confluent hypergeometric functions [13], respectively. The two fundamental solutions here are chosen such that each of them stands for a separate wave moving in a certain direction. This is shown in Fig.(2), where the three possible motion scenarios are presented by separate curves for the probability density $p(x) = |\psi(x)|^2$. It is seen that the case of a particle moving from the left to the right is described by the term proportional to ${}_1F_1$ ($c_2 = 0$), $U$-term stands for a particle moving from the right to the left ($c_1 = 0$), and the case $c_1 c_2 \neq 0$ describes the mixed situation of counter-propagating fluxes.

Let a particle move from the left to the right. The asymptote of the wave function at the positive infinity is

$$\psi(x) = (is)^{1-i\delta} \left( c_1(1-i\delta) - c_2 \frac{\Gamma(i\delta)}{\Gamma(ia)} \right) e^{\frac{i\delta x}{2\sigma}} + c_2 \frac{\Gamma(-i\delta)\left(as - \delta(\delta+s)/2\right)}{\Gamma(ia-i\delta+1)} e^{-\frac{i\delta x}{2\sigma}}, \qquad (7)$$

from which we see that to have an outgoing transmitted wave alone we should put $c_2 = 0$:

$$\psi(+\infty) \sim c_1 (is)^{1-i\delta} (1-i\delta) e^{i\delta x/2\sigma}. \qquad (8)$$

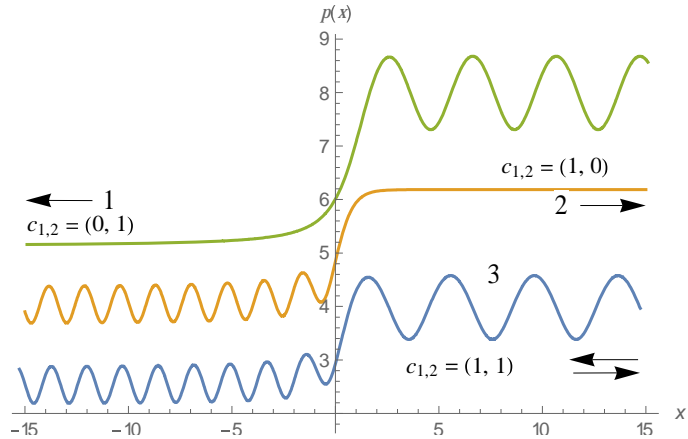

Fig. 2. Motion scenarios at above-barrier transmission of a quantum particle.
1) $c_{1,2} = (1,0)$ - from left to right, 2) $c_{1,2} = (0,1)$ - from right to left, 3) $c_{1,2} = (1,1)$ - mixed case of counter-propagating fluxes.

At the negative infinity $x \to -\infty$ the asymptote of the wave function presents a superposition of the incident and reflected waves:

$$\psi(x) = \frac{c_1}{2} i^{1-2i\delta} \Gamma(2-i\delta) z^{\frac{i\delta}{2}} (-isz)^{-i(a+\delta)} \left( \frac{(s-\delta)e^{isz/2}(sz)^{2ia}}{\Gamma(ia-i\delta+1)} + \frac{(s+\delta)e^{-isz/2}(-isz)^{i\delta}}{\Gamma(1-ia)} \right), \quad (9)$$

and the asymptote for the coordinate transformation $z(x)$ is as follows:

$$z(-\infty) \sim -\frac{x}{\sigma} + \log\left(-\frac{\sigma}{x}\right). \quad (10)$$

For the negative infinity, since $x < 0$, the reflected wave is given by the first term in equation (9). Consequently, for the reflection coefficient we get the result

$$R = \frac{a e^{-\pi\delta}}{a-\delta} \frac{(s-\delta)^2}{(s+\delta)^2} \frac{\sinh(\pi(a-\delta))}{\sinh(\pi a)} \quad (11)$$

or

$$R = e^{-2\pi\sigma k_2} \frac{\sinh\left(\frac{\pi\sigma}{2k_1}(k_1-k_2)^2\right)}{\sinh\left(\frac{\pi\sigma}{2k_1}(k_1+k_2)^2\right)}, \quad (12)$$

where the standard notations for the wave numbers are introduced:

$$k_1 = \sqrt{\frac{2mE}{\hbar^2}}, \quad k_2 = \sqrt{\frac{2m(E-V_0)}{\hbar^2}}. \quad (13)$$

It seems useful to compare this result with those for the abrupt-step and hyperbolic-tangent potential barriers:

$$V_{SP} = \begin{cases} 0, & x < 0 \\ V_0, & x \geq 0 \end{cases} \quad \text{and} \quad V_{HT} = \frac{V_0}{1+e^{-x/d}}, \quad (14)$$

for which the reflection coefficients read

$$R_{SP} = \frac{(k_1-k_2)^2}{(k_1+k_2)^2} \quad \text{and} \quad R_{HT} = \frac{\sinh^2(\pi d(k_1-k_2))}{\sinh^2(\pi d(k_1+k_2))}. \quad (15)$$

The graphs for the three reflection coefficients are shown in Fig.3.

Consider the limits of small and large $\sigma$. At $\sigma \to 0$ the exponent in equation (12) tends to the unity and the sinh functions become approximately equal to their arguments, so that in this limit $R$ for the Lambert potential recovers the result for the step-barrier:

$$R\big|_{\sigma \to 0} = \frac{(k_1-k_2)^2}{(k_1+k_2)^2}(1-2\pi\sigma k_2) + O(\sigma^2) \quad (16)$$

As it is seen, the effect of the barrier asymmetry in this limit is of the first order of smallness.

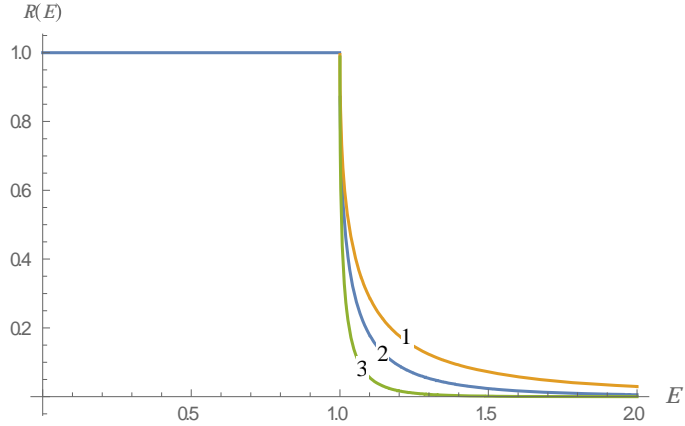

Fig.3. Comparison with the reflection $R_{SP}$ by the step - (**1**) and $R_{HT}$ by the hyperbolic-tangent - (**3**) potentials, $V_0 = 1$, $\sigma = 0.15$, $d = 0.5$.

For $\sigma \approx d$ the upper right-hand edge of the Lambert potential reproduces that of the hyperbolic tangent one (Fig.4). However, the left-hand side of the potential is of significantly lesser steepness (note that the fixed point for the Lambert potential is $V(0) \approx 0.638 V_0$ while for the hyperbolic tangent case $V(0) = V_0/2$). For this reason, the reflection becomes less than that of the hyperbolic tangent potential (inset of Fig.4).

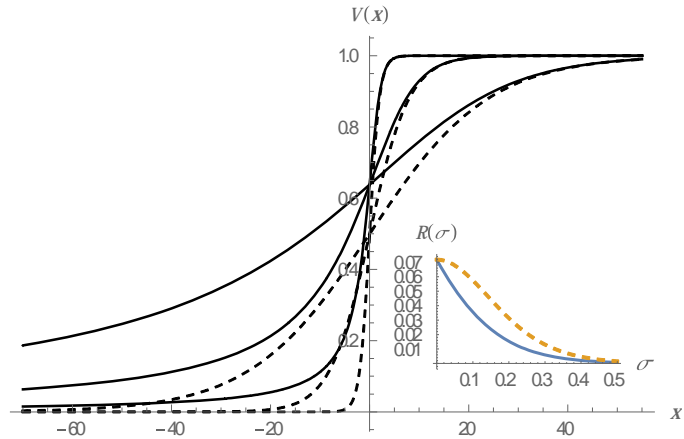

Fig.4. The Lambert-W (solid lines) and hyperbolic tangent (dashed lines) potentials, $V_0 = 1$, $\sigma = d = 1, 4, 12$. Inset presents the reflection coefficient versus $\sigma$ for $E = 1.5$ and $d = \sigma$.

## 3. Discussion

Thus, we have introduced an exactly solvable potential - the Lambert $W$-potential. This is an asymmetric step potential with controlled asymmetry and steepness. A distinct feature of the general solution of the Schrödinger equation for this potential is that it is written as a combination of four confluent hypergeometric functions with non-constant coefficients.

Discussing the properties of the potential, we have considered the quantum reflection at transmission of a particle above such a barrier. We have derived a compact formula for the reflection coefficient and have compared that with those for the abrupt-step and hyperbolic tangent potentials which are reproduced by our potential in certain parameter and/or variable variation regions. We have shown that the reflection coefficient is smaller because of the lesser steepness of the potential on the particle incidence side.

There are very few potentials for which the stationary Schrödinger equation is exactly solved in terms of special functions. We have recently presented a new confluent hypergeometric potential - the inverse square root potential [3]. Here we have introduced one more independent potential. The derivation of this potential employs an equation obeyed by a function involving the derivative of a solution of the bi-confluent Heun equation [6,7]. Since the approach that leads to the derivation of both the inverse square root potential and the Lambert product logarithm potential introduces new features in the search for exactly integrable models, we have presented the derivation lines specifically for the Lambert potential in a concluding Appendix A. We have shown that the result is in fact a more general five-parametric potential which is, however, a conditionally integrable one. The exactly solvable sub-potential is just the four-parametric Lambert $W$-potential that we have presented here. Finally, we have shown that there is one more potential for which the solution of the Schrödinger equation is written in terms of the derivative of a bi-confluent Heun function.

**Acknowledgments**

This research has been conducted within the scope of the International Associated Laboratory IRMAS (CNRS-France & SCS-Armenia). The work has been supported by the Armenian State Committee of Science (SCS Grant No. 13RB-052) and the project "Leading Russian Research Universities" (Grant No. FTI_120_2014 of the Tomsk Polytechnic University).

**Appendix A. Derivation of the Lambert-W potential**

We adopt here the following canonical form of the bi-confluent Heun equation:

$$\frac{d^2u}{dz^2} + \left(\frac{\gamma}{z} + \delta + \varepsilon z\right)\frac{du}{dz} + \frac{\alpha z - q}{z}u = 0. \tag{A1}$$

The integrability of the Schrödinger equation in terms of solutions of this equation has been considered by many authors (see, e.g., [14-21]). Lamieux and Bose gave a first systematic treatment of this question and have presented five six-parametric potentials [21]. We have recently shown that these are the only energy-independent exactly solvable potentials that are proportional to a continuous energy-independent parameter and have a shape that is independent of that parameter [5].

A further progress in constructing new solvable potentials is achieved if one considers the equations obeyed by the derivatives of the Heun functions. Recently, when discussing the case of the tri-confluent Heun function, we have arrived at the inverse square root potential [3]. Here we consider the equation obeyed by the following function involving the first derivative of the bi-confluent Heun function:

$$w = z^{\gamma} e^{\delta z + \varepsilon z^2/2} \frac{du}{dz}. \tag{A2}$$

The equation of interest deduced from the bi-confluent equation (A1) is written as

$$\frac{d^2w}{dz^2} - \left(\frac{\gamma - 1}{z} + \delta + \varepsilon z + \frac{1}{z - z_0}\right)\frac{dw}{dz} + \frac{\alpha(z - z_0)}{z}w = 0, \tag{A3}$$

where $z_0 = q/\alpha$. As it is seen, this equation possesses an additional singularity located at $z_0$.

Applying the transformation of the variables $\psi = \varphi(z)u(z)$, $z = z(x)$ to the Schrödinger equation, eliminating $\varphi(z)$ and matching the result with equation (A3), one gets

$$I(z) = g - \frac{f_z}{2} - \frac{f^2}{4} = \left[-\frac{1}{2}\left(\frac{\rho_z}{\rho}\right)_z - \frac{1}{4}\left(\frac{\rho_z}{\rho}\right)^2\right] + \frac{2m}{\hbar^2}\frac{E - V(z)}{\rho^2}, \tag{A4}$$

where $I(z)$ is the invariant, $f$ and $g$ are the coefficients of the $w'$ and $w$ terms of equation (A3), respectively, and $\rho = dz/dx$. According to the approach proposed in [5], the energy-independent potentials, which are proportional to an energy-independent parameter, are derived if the $\rho$- term in the square brackets, the energy term $E/\rho^2$, and the potential term $V(x)/\rho^2$ are *separately* matched with the invariant $I(z)$. Then, since the finite singularities of the invariant are $z=0$ and $z=z_0$, the coordinate transformation should be of the form $\rho = z'(x) = z^{m_1}(z-z_0)^{m_2}/\sigma$ with integer or half-integer $m_{1,2}$.

Matching the $\rho$- term with the corresponding term of the invariant we get $m_2 = -1$. Furthermore, since the invariant of equation (A3) is a sixth-degree polynomial in $z$ divided by $z^2(z-z_0)^2$, matching the energy term we get three possible choices for $m_1$: $m_1 = 0, 1/2, 1$. A further inspection shows that the first choice does not produce a solution written in terms of the hypergeometric functions, the second choice is not possible because of contradicting equations, so that we are left with $m_1 = 1$. Thus, we consider the coordinate transformation

$$\rho(z) = \frac{dz}{dx} = \frac{z/\sigma}{z-z_0}. \tag{A5}$$

Since the bi-confluent Heun equation is invariant with respect to the scaling transformation $z \to sz$, for convenience, without loss of generality, we put $z_0 = -1$ and change $\sigma \to -\sigma$. The result is the following Lambert product logarithm function:

$$z = W(e^{-(x-x_0)/\sigma}), \quad x_0 = \text{const}. \tag{A6}$$

Examining now the potential term in equation (A4) we see that the potential $V(z)$ is a sixth-degree polynomial divided by $(z-z_0)^4$. With this $V(z)$ rewritten as

$$V(z) = \frac{V_4}{(1+z)^4} + \frac{V_3}{(1+z)^3} + \frac{V_2}{(1+z)^2} + \frac{V_1}{1+z} + V_0 + V_5(1+z) + V_6(1+z)^2 \tag{A7}$$

and $\rho(z)$ given by equation (A5), equation (A4) is reduced to an over-determined set of seven algebraic equations for the four remaining free parameters involved in the bi-confluent Heun equation: $\gamma, \delta, \varepsilon, \alpha$ (we recall that we have put $z_0 = -1$ so that we have specified $q = -\alpha$). It is understood that these equations may have non-zero solutions only under rather severe restrictions imposed on the parameters of the potential (A7). The examination readily shows that it should be $V_4 = V_5 = V_6 = 0$ and

$$V_2 + V_3 = \frac{2m}{\hbar^2}\sigma^2 V_3^2. \tag{A8}$$

We thus arrive at a five-parametric *conditionally* integrable potential given as

$$V(z) = V_0 + \frac{V_1}{1+z} + \frac{-V_3 + 2m\sigma^2 V_3^2/\hbar^2}{(1+z)^2} + \frac{V_3}{(1+z)^3}, \quad z = W(e^{-(x-x_0)/\sigma}). \tag{A9}$$

The Lambert product logarithm potential (2) is the mass-independent four-parametric sub-potential of this, more general, potential achieved by the specification $V_3 = 0$ (in equation (2), we omitted two of the parameters, the energy origin and the space origin $x_0$). We note that the potential (A9) itself is a particular specification of a more general seven-parametric confluent Heun potential that we have recently presented in [5].

For the completeness of the treatment, we recall that the pre-factor $\varphi(z)$ is given as

$$\varphi(z) = \rho(z)^{-1/2} e^{\frac{1}{2}\int f(z)dz}, \qquad (A10)$$

hence, the solution of the Schrödinger equation is finally written as

$$\psi(z) = z^{\gamma/2} e^{(\delta z + \varepsilon z^2/2)/2} \frac{du(z)}{dz}, \qquad (A11)$$

where $u(z) = H_B(\gamma, \delta, \varepsilon, \alpha, q)$ is the solution of the starting bi-confluent Heun equation (A1), and the involved parameters are given as

$$\gamma = 2\sigma \sqrt{\frac{2m}{\hbar^2}\left(-E + V_0 + V_1 + \frac{2m}{\hbar^2}\sigma^2 V_3^2\right)}, \quad \delta = \gamma + \frac{4m\sigma^2 V_3}{\hbar^2}, \quad \alpha = -q = \frac{2m\sigma^2(V_1 + \delta V_3)}{\hbar^2} \qquad (A12)$$

and $\varepsilon = 0$. The last equation accomplishes the development because for $\varepsilon = 0$ the bi-confluent Heun function is reduced to a product of an elementary function and a confluent hypergeometric function.

We conclude by noting that we have not discussed here one of the cases when the Schrödinger equation is also solved in terms of the derivative of a bi-confluent Heun function, namely, the case corresponding to the choice $m_{1,2} = (0, -1)$ in the coordinate transformation $z'(x) = z^{m_1}(z - z_0)^{m_2}/\sigma$. The corresponding potential reads

$$V(x) = V_0 + \frac{V_1}{\sqrt{x}\left(\sqrt{x} + z_0\right)}. \qquad (A13)$$

Though the solution of the Schrödinger equation in this case is not written in terms of simpler special functions, however, the solution possesses specific analytic properties resulting in distinct physical behaviour. We hope to treat this potential on a different occasion.